\documentclass[12pt,a4paper]{article}

\title{Box model of migration in channels of migration networks}

\author{Nikolay K. Vitanov, Kaloyan N. Vitanov, Tsvetelina Ivanova}
\date{Institute of Mechanics, Bulgarian Academy of Sciences, Acad. G. Bonchev Str., Bl. 4, 1113 Sofia, Bulgaria}

\begin{document}
\maketitle

\abstract{We discuss a box model of migration in channels of networks with possible
application for modelling motion of migrants in migration networks.
The channel consists of nodes of the network (nodes may be considered
as boxes representing countries) and edges that connect these nodes and
represent  possible ways for motion of migrants. The nodes  of the migration
channel have different "leakage", i.e. the probability of change of the
status of a migrant (from migrant to non-migrant) may be different in the
different countries along the channel. In addition the nodes far from the
entry node of the channel may be more attractive for migrants in comparison to
the nodes around the entry node of the channel. We discuss below channels containing 
infinite number of nodes. Two regimes of functioning of these channels are studied: stationary regime and non-stationary regime. In the stationary regime of the functioning of the channel the distribution of migrants in the countries of the channel is described by a distribution that contains as particular case the Waring distribution. In the non-stationary regime of functioning of the channel one observes exponential increase or exponential decrease of the number of migrants in the countries of the channel. It depends on the situation in the entry country of the channel for which scenario will be realized. Despite the non-stationary regime of the functioning of the channel the asymptotic distribution of the migrants in the nodes of the channel is stationary. From the point of view of the characteristic features of the migrants we discuss the cases of (i) migrants
having the same characteristics and (ii) two classes of migrants that have
differences in some characteristic (e.g., different religions).
}

\section{Introduction}
Flows in complex networks are  important for existence and functioning of the
systems containing such networks. Human migration (the permanent or
semipermanent change of residence that involves e.g., the relocation of
individuals, households or  moving groups between geographical locations \cite{everet}) 
is  one example of such a flow \cite{fawcet}, \cite{gurak}. . 
Large external migration  flows reached  Europe in  the last
years and this makes the study of migration very actual topic. In addition the
internal migration studies are important for taking decisions about economic
development of regions of a country \cite{armi}-\cite{champi}, \cite{eti}, \cite{grd05}, \cite{ht},  \cite{smn}, \cite{skeldon}, \cite{vit9}.  
Examples for results from such studies is e.g., the Hecksehr-Ohlin  theorem for economic use
of a country relative abundant factors such as labor  as well as the
factor-price equalization theorem \cite{borj}.
Migrant flows may be modelled by deterministic or
stochastic tools \cite{hott}, \cite{led90}, \cite{massey}, \cite{puu}, \cite{puu2}, \cite{whaag} and  the corresponding
migration models can be classified as probability models \cite{will99},
\cite{will08} or deterministic models with respect to their
mathematical features. Examples for probability models are the
exponential model, multinomial model or Markov chain models of migration.
One of the most famous deterministic models of migration is
the gravity model of migration \cite{grd05}. The gravity model may be
extended in different ways, e.g., to include the income and unemployment in
the two regions. 
\par
Human migration is closely connected to ideological struggles \cite{vit1},
\cite{vit2} and waves and statistical distributions in population systems \cite{vz1},
\cite{vit6}, \cite{vit3}, \cite{vit4}, \cite{vit5}. In this article we shall consider a box model
of a flow of migrants in a channel (sequence of countries) of a migration
network. The nodes (countries) will be considered as boxes (cells)  where the following processes happen: 
inflow and outflow of migrants and "leakage" (change of the status of migrant). Migrants enter the channel from the entry country and
move through the channel.
The different nodes of the channel (the different countries) are assumed to have different rate of "leakage" 
(i.e. different probabilities of change of the status of a person form the migrant to non-migrant). 
\par
The paper is organized as follows. In Sect.2 the model for moving of
substance in a channel  containing an infinite number of nodes is discussed. Two regimes of functioning  of the channel: stationary regime (the amount of the
substance in the entry  box of the channel doesn't change) and non-stationary regime (amount of substance in the entry box of the channel decreases
or increases exponentially) are described. Statistical distributions of the
amount of substance in the nodes of the channel are obtained. A particular case
of the distribution for the stationary regime of functioning of the channel is
the Waring distribution. Sect. 3 is devoted to the case of two immiscible substances
moving through the channel channel. In Sect. 4 we relate the obtained mathematical results
to the movement of migrants through a migration channel.
\section{Channel containing infinite number of nodes}
Inspired by the models in \cite{sg1}, \cite{vk1}, and \cite{vv1} we consider a  model of moving of a substance through a channel as follows.  The channel contains infinite number
of nodes and each node can be considered as  a cell. The cells are indexed in  succession by non-negative integers. The first cell
has index $0$. We assume that an amount $x$ of some substance  is
distributed among the cells and this substance can move from one cell to
another cell. Let $x_i$ be the amount of the substance in the $i$-th cell.
Then
\begin{equation}\label{warigx1}
x = \sum \limits_{i=0}^\infty x_ i
\end{equation}
The fractions $y_i = x_i/x$ can be considered as probability values of
distribution of a discrete random variable $\zeta$
\begin{equation}\label{warigx2}
y_i = p(\zeta = i), \ i=0,1, \dots
\end{equation}
The content $x_i$ of any cell can change because of the following 3 processes:
\begin{enumerate}
\item Some amount $s$ of the substance $x$  enters the channel
from the external environment through the $0$-th cell;
\item Rate $f_i$ from $x_i$ is transferred from the $i$-th
cell into the $i+1$-th cell;
\item Rate $g_i$ from  $x_i$  leaks out the $i$-th cell into the
external environment.
\end{enumerate}
The above processes can be modeled mathematically by the system of ordinary
differential equations:
\begin{eqnarray} \label{warigx4}
\frac{dx_0}{dt} &=& s-f_0-g_0; \nonumber \\
\frac{dx_i}{dt} &=& f_{i-1} -f_i - g_i, \ i=1,2,\dots.
\end{eqnarray}
The following forms of the amount of the moving substances may be assumed
($\alpha, \beta, \gamma_i, \sigma$ are constants)
\begin{eqnarray}\label{warigx5}
s &=& \sigma x_0; \ \ \sigma > 0  \nonumber \\
f_i &=& (\alpha + \beta i) x_i; \ \ \ \alpha >0, \ \beta \ge 0 \to
\textrm{cumulative advantage of higher cells} \nonumber \\
g_i &=& \gamma_i x_i; \ \ \ \gamma_i \ge 0 \to \textrm{non-uniform leakage 
over the cells}
\end{eqnarray}
The rules (\ref{warigx5}) differ from the rules in \cite{sg1} as follows:
\begin{enumerate}
\item $s$ is proportional to the the amount of the substance $x_0$ in the $0$-th node. In \cite{sg1} $s$ is proportional to the amount $x$ of the substance
in the entire channel;
\item Leakage rates $\gamma_i$ are different for the different nodes. In
\cite{sg1} and \cite{vv1} the leakage rate is constant and equal to $\gamma$ for all nodes
of the channel (i.e., there is uniform leakage over the cells).
\end{enumerate}
\par
Substitution of Eqs.(\ref{warigx5}) in Eqs.(\ref{warigx4}) leads to the
relationships
\begin{eqnarray}\label{wsx10}
\frac{dx_0}{dt} &=& \sigma x_0 - \alpha x_0 - \gamma_0 x_0; \nonumber \\
\frac{dx_i}{dt} &=& [\alpha+ \beta(i-1)]x_{i-1} - (\alpha + \beta i + \gamma_i)x_i; \ \ 
\ i=1,2,\dots
\end{eqnarray}
There are  two regimes of functioning of the channel and realization of one
of them depends on the situation in the $0$-th node (the entry cell). The regimes are  stationary regime and non-stationary regime.
\subsection{Stationary regime of functioning of the channel}
In the stationary regime of the functioning of the channel $\sigma = \alpha + \gamma_0$ which means that $x_0$
(the amount of the substance in the $0$-th cell of the channel) is free
parameter. In this case the solution of Eqs.(\ref{wsx10}) is
\begin{equation}\label{wsx11}
x_i = x_i^* + \sum \limits_{j=0}^i b_{ij} \exp[-(\alpha + \beta j + \gamma_j)t]
\end{equation}
where $x_i^*$ is the stationary part of the solution. For $x_i^*$ one obtains
the relationship
\begin{equation}\label{wsx12}
x_i^* = \frac{\alpha + \beta (i-1)}{\alpha + \beta i + \gamma_i} x_{i-1}^*
\end{equation}
The corresponding relationships for the coefficients $b_{ij}$ are
\begin{equation}\label{wsx13}
b_{ij} = \frac{\alpha + \beta(i-1)}{\gamma_i - \gamma_j + \beta(i-j)} b_{i-1,j},
\ j=0,1,\dots,i-1
\end{equation}
From Eq.(\ref{wsx12}) one obtains
\begin{equation}\label{wsx14}
x_i^* = \frac{[k+(i-1)]!}{(k-1)! \prod \limits_{j=1}^i (k+j+a_j)} x_0^*
\end{equation}
where $k = \alpha/\beta$ and $a_j=\gamma_j/\beta$.
The form of the corresponding stationary distribution $y_i^* = x_i^*/x^*$ (where $x^*$ is the amount of the substance in all of the cells of the channel)
is
\begin{equation}\label{wsx15}
y_i^* = \frac{[k+(i-1)]!}{(k-1)! \prod \limits_{j=1}^i (k+j+a_j)} y_0^*
\end{equation}
Let us consider the particular case where $a_0 = a_1 = \dots = a$. In this case the
distribution from Eq.(\ref{wsx15}) is reduced to the distribution:
\begin{eqnarray}\label{wsx16}
P(\zeta = i) &=& P(\zeta=0) \frac{(k-1)^{[i]}}{(a+k)^{[i]}}; \ \ k^{[i]} = \frac{(k+i)!}{k!}; \ i=1, 2, \dots 
\end{eqnarray}
$P(\zeta=0)=y_0^* = x_0^*/x^*$ is the percentage of substance that is located in
the first cell of the channel. Let this percentage be
\begin{equation}\label{wsx17}
y_0^* = \frac{a}{a+k}
\end{equation}
The case described by Eq.(\ref{wsx16}) corresponds to the situation where the
amount of substance in the first cell is proportional of the amount of substance
in the entire channel (self-reproduction property of the substance). In this
case Eq.(\ref{wsx15}) is reduced to the  distribution:
\begin{eqnarray}\label{wsx18}
P(\zeta = i) &=& \frac{a}{a+k} \frac{(k-1)^{[i]}}{(a+k)^{[i]}}; \ \ k^{[i]} = \frac{(k+i)!}{k!}; \ i=1, 2, \dots 
\end{eqnarray}
Let us denote $\rho = a$ and $k=l$. The distribution (\ref{wsx18}) is
exactly the Waring distribution (probability distribution of non-negative integers named after Edward Waring - a Lucasian professor 
of Mathematics in Cambridge in the 18th century)  \cite{varyu3}, \cite{varyu1}, \cite{varyu2}
\begin{equation}\label{ap1}
p_l = \rho \frac{\alpha_{(l)}}{(\rho + \alpha)_{(l+1)}}; \
\alpha_{(l)} = \alpha (\alpha+1) \dots (\alpha+l-1)
\end{equation}
Waring distribution may be written also as follows
\begin{eqnarray}\label{ap2}
p_0 &=& \rho \frac{\alpha_{(0)}}{(\rho + \alpha)_{(1)}} = \frac{\rho}{\alpha + \rho}
\nonumber \\
p_l &=& \frac{\alpha+(l-1)}{\alpha+ \rho + l}p_{l-1}.
\end{eqnarray}
The mean $\mu$ (the expected value) of the Waring distribution is
\begin{equation}\label{ap3}
\mu = \frac{\alpha}{\rho -1} \ \textrm{if} \ \rho >1
\end{equation}
The variance of the Waring distribution is
\begin{equation}\label{ap4}
V = \frac{\alpha \rho (\alpha + \rho -1)}{(\rho-1)^2(\rho - 2)} \
\textrm{if} \ \rho >2
\end{equation}
$\rho$ is called the tail parameter as it controls the tail of the Waring
distribution. Waring distribution contains various distributions as particular cases. Let $i \to \infty$ Then the Waring distribution is reduced to
\begin{equation}\label{ap5}
p_l \approx \frac{1}{l^{(1+\rho)}}.
\end{equation}
which is the frequency form of the Zipf distribution \cite{chen}.
If $\alpha \to 0$ the Waring distribution is reduced to the Yule-Simon distribution \cite{simon}
\begin{equation}\label{ap6}
p(\zeta = l \mid \zeta > 0) = \rho B(\rho+1,l)
\end{equation}
where $B$ is the beta-function.
\subsection{Non-stationary regime of functioning of the channel}
In the nonstationary case $dx_0/dt \ne 0$. In this case the solution of the
first equation of the system of equations (\ref{wsx10}) is
\begin{eqnarray}\label{nst11}
x_0 = b_{00} \exp[(\sigma - \alpha - \gamma_0)t] 
\end{eqnarray}
where $b_{00}$ is a constant of integration. $x_i$ must be obtained by solution of the corresponding  Eqs.(\ref{wsx10}). The form of $x_i$ is
\begin{eqnarray}\label{nst12}
x_i = \sum \limits_{j=0}^i b_{ij} \exp[-(\alpha + \beta j + \gamma_j - \sigma_j )t]
\end{eqnarray}
The solution of the system of equations (\ref{wsx10}) is (\ref{nst12})
where $\sigma_i =0$, $i=1,\dots,$:
\begin{eqnarray}\label{nst13}
b_{ij} &=& \frac{\alpha + \beta(i-1)}{\gamma_i - \gamma_j + \beta(i-j)} b_{i-1,j}; \ i=1,\dots, 
\end{eqnarray}
and $b_{ii}$ are determined from the initial conditions in the cells of the
channel. The asymptotic solution ($t \to \infty$) is
\begin{equation}\label{nst13}
x_i^a = b_{i0} \exp[(\sigma - \alpha - \gamma_0)t]
\end{equation}
This means that the asymptotic distribution $y_i^a = x_i^a/x^a$ is stationary
\begin{equation}\label{nst14}
y_i^a = \frac{b_{i0}}{\sum \limits_{j=0}^\infty b_{j0}}
\end{equation}
regardless of the fact that the amount of substance in the two cells may increase or decrease exponentially. The explicit form of this distribution is
\begin{eqnarray}\label{nst15}
y_0^a = \frac{1}{1+ \sum \limits_{i=1}^{\infty} \prod \limits_{k=1}^i
\frac{\alpha + \beta(k-1)}{\gamma_k - \gamma_0 + \beta k}},   \ \
y_i^a = \frac{\prod \limits_{k=1}^i \frac{\alpha + \beta (k-1)}{\gamma_k - \gamma_0 + \beta k}}{\sum \limits_{i=0}^{\infty} \prod \limits_{k=1}^i
\frac{\alpha + \beta(k-1)}{\gamma_k - \gamma_0 + \beta k}}, 
i=1,\dots
\end{eqnarray}
\section{The model of two substances}
Let us  discuss now a model of moving of two immiscible substances through a channel
containing infinite number of cells. The substances enter the channel through the
entry cell and in general the following three processes are allowed: the
substances may enter the cells one after the another and the substances may
be used for some purposes in the corresponding cell. From the point of view
of migration flows this model corresponds to migration of migrants with two
different values of some characteristics (e.g. different religions). 
\par
Let us denote the amount of substance of the two types in the $i$-th cell of
the channel as $x^1_i$ and $x^2_i$. The model equations for the movement of the
two kings of substance are
\begin{eqnarray}\label{2s1}
\frac{dx^1_0}{dt} &=& \sigma^1 x^1_0 - \alpha^1 x^1_0 - \gamma^1_0 x^1_0; \nonumber \\
\frac{dx^1_i}{dt} &=& [\alpha^1 + \beta^1 (i-1)]x^1_{i-1} - (\alpha^1 + \beta^1 i + \gamma^1_i)x^1_i; \ \ 
\ i=1,2,\dots
\end{eqnarray}

\begin{eqnarray}\label{2s2}
\frac{dx^2_0}{dt} &=& \sigma^2 x^2_0 - \alpha^2 x^2_0 - \gamma^2_0 x^2_0; \nonumber \\
\frac{dx^2_i}{dt} &=& [\alpha^2 + \beta^2(i-1)]x^2_{i-1} - (\alpha^2 + \beta^2 i + \gamma^2_i)x^2_i; \ \ 
\ i=1,2,\dots
\end{eqnarray}
For the stationary regime of functioning of the channel the amount of the substances and the
stationary distributions of the substances in for the two kinds of substances are
\begin{equation}\label{2s3}
x_i^{1,*} = \frac{[k^1+(i-1)]!}{(k^1-1)! \prod \limits_{j=1}^i (k^1+j+a^1_j)} x_0^{1,*}
\end{equation}
\begin{equation}\label{2s4}
y_i^{1,*} = \frac{[k^1+(i-1)]!}{(k^1-1)! \prod \limits_{j=1}^i (k^1+j+a^1_j)} y_0^{1,*}
\end{equation}
\begin{equation}\label{2s5}
x_i^{2,*} = \frac{[k^2+(i-1)]!}{(k^2-1)! \prod \limits_{j=1}^i (k^2+j+a^2_j)} x_0^{2,*}
\end{equation}
\begin{equation}\label{2s6}
y_i^{2,*} = \frac{[k^2+(i-1)]!}{(k^2-1)! \prod \limits_{j=1}^i (k^2+j+a^2_j)} y_0^{2,*}
\end{equation}
where $k^1=\alpha^1/\beta^1$; $k^2=\alpha^2/\beta^2$; $a_j^1 = \gamma_j^1/\beta^1$;
 $a_j^2 = \gamma_j^2/\beta^2$; $y_i^{1,*}=x_i^{1,*}/x^{1,*}$; $y_i^{2,*}=x_i^{2,*}/x^{2,*}$
 and $x^{1,*}$ and $x^{2,*}$ are the total amounts of the two substances in all cells of the
 channel. 
 \par 
 For the case of non-stationary regime of functioning of the channel the forms of the asymptotic 
 distribution for the two kinds of substances are
 \begin{equation}\label{2s7}
 y_0^{1,a} = \frac{1}{1+ \sum \limits_{i=1}^{\infty} \prod \limits_{k=1}^i
\frac{\alpha^1 + \beta^1 (k-1)}{\gamma^1_k - \gamma^1_0 + \beta^1 k}}, \ \ 
y_i^{1,a} = \frac{\prod \limits_{k=1}^i \frac{\alpha^1 + \beta^1 (k-1)}{\gamma^1_k - \gamma^1_0 + \beta^1 k}}{\sum \limits_{i=0}^{\infty} \prod \limits_{k=1}^i
\frac{\alpha^1 + \beta^1 (k-1)}{\gamma^1_k - \gamma^1_0 + \beta^1 k}}, 
i=1,\dots
 \end{equation}
 \begin{equation}\label{2s8}
 y_0^{2,a} = \frac{1}{1+ \sum \limits_{i=1}^{\infty} \prod \limits_{k=1}^i
\frac{\alpha^2 + \beta^2 (k-1)}{\gamma^2_k - \gamma^2_0 + \beta^2 k}},   \ \
y_i^{2,a} = \frac{\prod \limits_{k=1}^i \frac{\alpha^2 + \beta^2 (k-1)}{\gamma^2_k - \gamma^2_0 + \beta^2 k}}{\sum \limits_{i=0}^{\infty} \prod \limits_{k=1}^i
\frac{\alpha^2 + \beta^2 (k-1)}{\gamma^2_k - \gamma^2_0 + \beta^2 k}}, 
i=1,\dots
 \end{equation}

\section{Discussion}
We have mentioned above that the discussed model can be used for study of movement
and distribution of migrants in a sequence of countries that form a migration channel.
Migration channel of such kind was clearly visible in 2015 when a large influx of migrants
in Europe was observed along a channel with Greece as the entry country. The parameters of the models
discussed above can be interpreted as follows from the point of view of the 
application of the models to the migration channels. $\sigma$ can be considered as
a "gate" parameter as it regulates the number of migrants that enter the channel.
$\sigma$ is a parameter that is specific for the entry country of the channel.
A small value of $\sigma$ can decrease significantly the number of migrants that
enter the channel. Large value of $\sigma$ may lead to large migration flows.
The value of the parameter $\sigma$ can be regulated by the authorities of the 
entry country and by some over-national authorities if such authorities exist and
they are allowed to act on the territory of the entry country. Thus if the state
structures of the entry country are weak because of some kind of crisis or as a
consequences of other reasons then the value of the parameter $\sigma$ may be
large. Thus large number of migrants may enter the channel and this will lead
to large problems in the entry country and in all countries along the channel, especially
to the countries that are close the entry country and are part of the migration channel.
\par 
Another kind of "gate" parameter is the parameter $\alpha$ that regulates the number
of migrants that move from one country to the next country in the sequence of
countries that form the migration channel. Large value of $\alpha$ means that the 
movement between the countries is large and the migrants easily cross the state borders.
Small value of $\alpha$ means that the the crossing of the borders is more difficult.
What was observed for the case of the migration flows in 2015 in Europe was that at
some moments the borders have been practically open and for some time the value of the
parameter $\alpha$ was large and almost the same for all countries that have been part
of the channel. Such kind of situation is modelled by the models presented above.
\par 
The parameter $\beta$ accounts for the attractiveness of the countries of the channel that
are distant from the entry country of the channel. The large values of this parameter
lead to a tendency for leaving the countries around the beginning of the migration
channel and attempts to settle in much more attractive countries along the channel. 
Parameter $\gamma_i$ accounts for the number of migrants that enter the $i$-th
country of the channel but do not leave it. The reasons for this may be different, e.g.
some migrants may obtain permission to stay in the country. Small values of the parameters
$\gamma_i$ correspond to a large traffic of migrants through the countries of the channel.
If in some country the value of the corresponding parameter $\gamma$ is large then
significant number of migrants may stay in this country and the number of the 
migrant moving further through the channel may decrease.
\par 
For the case of two kinds of migrants the number of parameters increase which leads to 
increasing number of possible situations. For an example the entry country may prefer 
one of the kinds of migrants and then the values of the "gate" parameters $g^1$ and $g^2$
may have quite different values. The countries along the channel may impose different level of
difficulty of crossing the borders which may lead in differences in the values of the
parameters $\alpha^1$ and $\alpha^2$. The countries along the channel may have different 
level of attractiveness which will lead to different values of the parameters $\beta^1$ and
$\beta^2$. Finally different countries may have different preferences about the numbers and
about the kind of migrants they allow to stay in the country. This will lead to different values
of the parameters $\gamma^1_i$ and $\gamma^2_i$. 
\par
Let us discuss the above remarks in more detail by means of the obtained mathematical results.
Let us first consider the case of one kind of migrants and stationary regime of functioning of
the channel. From Eq.(\ref{wsx15}) we obtain 
\begin{equation}\label{d1}
\frac{y_i^*}{y_{i+1}^*} = 1 + \frac{\beta + \gamma_{i+1}}{\alpha + \beta i}
\end{equation}
Eq.(\ref{d1}) shows us that for the case of infinite channel the number of migrants 
has to decrease with increasing value of $i$, i.e., in the countries that are away from 
the entry country of the channel the number of migrants in any country is smaller than the 
number of migrants in the previous country of the channel. Then for the case of infinite channel 
the effect of concentration of migrants in countries that are far from the entry country is not 
observed even if these countries are very attractive to the migrants.
Let us note here that the models considered above can be extended to the
case of a channel containing finite number of cells (finite number of countries
for the vase of migrant flows). Then a new effect will be observed: concentration
of migrants in the last cell of the channel (the final destination country). If the
attractiveness of the final destination country is large then the concentration of
the migrants will be observed in the entry country of the channel and in the final
destination country. Such possible developments of the situation in the migration 
channels will be discussed elsewhere.
\par 
Eq.(\ref{d1}) shows additional details  about the influence of the parameters of the 
channel on the migrant flow. The increase of $\alpha$ leads to decreasing of the 
ratio $y_i^*/y_{i+1}^*$. This means that the larger value of the gate parameter $\alpha$ leads to 
a smoothing of the distribution of the number of migrants along the countries of the channel.
Thus if the countries that are far from the entry country of the channel want to 
decrease the number of migrants on its own territory the have to take measures to 
decrease the value of the parameter $\alpha$. The increasing of the value of the 
parameter $\beta$ has more complicated influence on the ratio $y_i^*/y_{i+1}^*$
but in principle the effect is the same as the increasing of the value of parameter $\alpha$.
If the countries that are attractive for migrants have a policy to decrease their
attractiveness then smaller number of migrants will reach their territory. If such
countries have "open doors" politics towards migrants then larger number of migrants will
leave the countries around the entry country of the channel and will move towards the 
attractive countries. In combination with large value of $\sigma$ this may lead to
floods of migrants in the attractive countries. But this will lead also to even larger
flood of migrants in the countries that are around the entry country of the channel.
This was observed indeed for the case of the massive migration in Europe in 2015.
Finally the increasing value of the "leakage" parameter $\gamma_{i+1}$ will lead to
decreasing of the ratio $y_i^*/y_{i+1}^*$. The reason for this is obvious: when more
migrants change their migrant status (e.g. obtain permission to stay) in the the $i+1$-th
country then the number of migrants without status will decrease and this will lead to decreasing
ratio $y_i^*/y_{i+1}^*$.
\par 
Let us now discuss further the case of two kinds of migrants for the stationary regime of
the functioning of the channel. In this case the ratio of the numbers of migrants in a
cell of the channel is
\begin{equation}\label{d2}
\frac{x_i^{1,*}}{x_i^{2,*}} = \frac{x_0^{1,*}}{x_0^{2,*}} \frac{[k^1 +(i-1)]!}{{[k^2+(i-1)]}!}
\frac{(k^2-1)!}{(k^1-1)!} \frac{\prod \limits_{j=1}^i (k^2 + j + a_j^2)}{\prod \limits_{j=1}^i
(k^1 + j + a^1_j)}
\end{equation}
Eq.(\ref{d2}) leads to the following conclusions. First of all the mix of the migrants in any
country of the channel depends on the ratio $x_0^{1,*}/x_0^{2,*}$ at the entry country of the 
channel. Thus  the politics in the entry country of the channel towards the different categories
of migrants  is extremely important for the distribution of migrants in the entire channel. If there
are migrants with unfavorable characteristics from the point of view of countries from the 
the channel then one of most effective ways to reduce the number of such migrants is to reduce 
their possibility to enter the channel. The influence of the other parameters of the channel is
more complicated. Large values of $\gamma_i^2$ with respect to the values of $\gamma_i^1$
(this corresponds, e.g., for much larger acceptance of the migrants of class 2 in comparison to the
migrants of class 1 in all countries of the channel up to the $i$-th country of the channel) when
the other parameters of the channel are equal for the two classes of migrants lead to larger value of the 
ratio $x_0^{1,*}/x_0^{2,*}$. This means that the mix of migrants without status in the $i$-th
country of the channel depends on the politics  of previous countries of the channel towards different 
categories of migrants. Finally the ratio of the migrants depends on the parameters $k^1$ and $k^2$
that are ratios $\alpha^1/\beta^1$ and $\alpha^2/\beta^2$, i.e. the ratios between corresponding 
possibility for mobility of the class of migrants (the parameters $\alpha^1$ and $\alpha^2$) and the
attractiveness of the countries far away from the entry country of the channel (the parameters $\beta^1$ and
$\beta^2$).  This dependence is the most complicated one. If the ratios $k^1$ and $k^2$ are the same 
then the influence of $k^1$ and $k^2$ adds nothing to the influence of parameters $a^1$ and $a^2$
on the mix of migrants in the countries of the channel. In order to obtain more information about the 
influence of $k^{1,2}$ (respectively about the influence of the parameters $\alpha^{1,2}$ and $\beta^{1,2}$
we can use the recurrence relationships for the numbers of migrants in the $i$-th country of the 
channel. These relationships may be written as
\begin{equation}\label{d3}
x^{1,*}_i =\frac{1}{1+\frac{\beta^1 + \gamma_i^1}{\alpha^1 + \beta^1 (i-1)}} x^{1,*}_{i-1}, \ 
x^{2,*}_i = \frac{1}{1+\frac{\beta^2 + \gamma_i^2}{\alpha^2 + \beta^2 (i-1)}}x^{2,*}_{i-1}
\end{equation}
Then the vary large values of the parameter $\beta$  increase the traffic of migrants through the channel and
the decrease of the number of migrants in the $i$ in comparison to the number of migrants in $i-1$-th country will be
approximately $1-1/i$ as a proportion. The influence of the very large values of the parameters $\alpha$ is larger.
When the value of the parameter $\alpha$ is much large that the values of the corresponding parameters
$\beta$ and $\gamma_i$ then the number of migrants in the $i$-th country of the channel will be approximately 
equal to the number of migrants in the $i$-th country of the channel. Thus the decreasing permeability of the
borders between the neighbouring states from the migration channel is more effective that decreasing of the 
attractiveness of the countries far from the entry country of the channel. 
\par
Let us mention finally several methodological renarks. The model presented above
was a linear one. One possible extention is to take into account nonlinear effects. Then the model system will contain nonlinear ODEs 
or nonlinear PDEs. Such equations may be soved analytically in some particular cases 
\cite{d1} - \cite{d3},  \cite{vy1} - \cite{vy4}, \cite{vit7}, \cite{vit8}, but in the most cases one has to solve them
numerically. The corresponding statistical distrubutions must be obtained numerically too.

\end{document}